\documentclass[a4paper,12pt]{article}
\usepackage{amsmath}
\usepackage{amssymb}
\usepackage{tensor}
\usepackage{latexsym}
\usepackage{enumerate}
\usepackage{bbm}
\usepackage{amsthm}
\usepackage{hyperref}
\newcommand{\pound}{\emph{\textsterling}}
\newcommand{\be}{\begin{equation}}
\newcommand{\ee}{\end{equation}}
\newcommand{\ben}{\begin{equation*}}
\newcommand{\een}{\end{equation*}}
\newcommand{\bea}{\begin{eqnarray}}
\newcommand{\eea}{\end{eqnarray}}
\newcommand{\bean}{\begin{eqnarray*}}
\newcommand{\eean}{\end{eqnarray*}}

\newcommand{\bsub}{\begin{subequations}}
\newcommand{\esub}{\end{subequations}}

\newcommand{\disfrac}[1][2]{\displaystyle\frac}

\newcommand{\bbar}{\overline}

\begin{document}
\numberwithin{equation}{section}

\title{FLRW metric $f(R)$ cosmology with a perfect fluid by generating integrals of motion}
\vspace{1cm}
\author{\textbf{N. Dimakis}\thanks{nsdimakis@gmail.com}\,, \textbf{T. Christodoulakis}\thanks{tchris@phys.uoa.gr}\,,
\textbf{Petros A. Terzis}\thanks{pterzis@phys.uoa.gr}\\
{\it Nuclear and Particle Physics Section, Physics
Department,}\\{\it University of Athens, GR 157--71 Athens}}

\date{}
\maketitle
\begin{abstract}
In the context of metric $f(R)$ gravity, we consider a FLRW space-time, filled with a perfect fluid described by a barotropic equation of state ($p = \gamma \rho$). We give the equivalent mini-superspace description and use the reparametrization invariance of the resulting Lagrangian to work in the equivalent constant potential description. At that point, we restrict our analysis to those models for which the ensuing scaled mini-superspace is maximally symmetric. Those models exhibit the maximum number of autonomous integrals of motion linear in the momenta, which are constructed by the Killing vectors of the respective mini-supermetric. The integrals of motion are used to analytically solve the equations of the corresponding models. Finally, a brief description of the properties of the resulting Hubble parameters is given.
\end{abstract}

\section{Introduction}

General relativity, since its formulation, led to many experimental
predictions. In the field of cosmology it provided us with the
Standard Model, which can be considered quite successful with
respect to a series of observational facts. Of course there exist
some discrepancies as well, such as flatness and/or horizon problems
together with the observed accelerated expansion of the universe at
present times. In order to address these issues, many deviations
from the gravitational theory of General Relativity have been
proposed. The most widespread among them is the theory of $f(R)$
gravity, where the scalar curvature $R$ in the usual Einstein
Hilbert action is replaced by a non linear function of itself. The
method by which the field equations of motion are obtained defines
different kinds of theories: a) metric, b) affine and c)
metric-affine $f(R)$ gravity (for an overview see \cite{Sot} and
references therein).

In cosmology, there exists an increasing interest in the study of
analytical solutions for a variety of $f(R)$ models (\cite{Carloni},
\cite{Clifton}, \cite{Shamir}, \cite{Domaz}) particularly by the use
of Noether symmetries (\cite{Cap1}, \cite{Bas}, \cite{Andr},
\cite{Cap2}), even at the quantum regime \cite{Vakili}. There also
have been proposed criteria for testing whether particular models
are physically accepted or not (\cite{Dolg}, \cite{Wang}), although
these are not always dogmatically imposed. Nevertheless, and in
spite of the physical significance, one can not overlook the
mathematical interest in the procedure of obtaining analytical
solutions, in particular wherever symmetry conditions are imposed.
In this paper, we adopt the metric $f(R)$ gravity point of view,
i.e. we consider the set of equations of motion that is obtained by
varying the action with respect to the spacetime metric
$g_{\mu\nu}$. We consider an FLRW spacetime and supply the model
with a perfect fluid satisfying the barotropic equation $p = \gamma
\rho$, where $p$ is the pressure, $\rho$ the energy density and
$\gamma$ a constant. Step by step, we construct the equivalent
mini-superspace model having as configuration variables the scalar
curvature $R$ and the scale factor $a$. The Euler - Lagrange
equations for this system are seen to be equivalent to the field
equations of the model under consideration and, additionally,
consistent with the definition of the scalar curvature. By using the
reparameterization invariance of the theory, we scale the lapse
function in order to work in the constant potential
parameterization. At this point the Killing vector fields of the
scaled mini-superspace metric are used to define autonomous
integrals of motion linear in the momenta. We choose to study those
particular $f(R)$ models for which the ensuing mini-superspace
exhibits the maximum number of symmetries, meaning that the
mini-supermetric is maximally symmetric. The corresponding integrals
of motion are used in order to completely integrate the equations of
motion and find the general analytical solution in each case.

The paper is organized as follows: In section \ref{sect1} we give a
brief outline of the theory regarding the existence and derivation
of linear integrals of motion for \emph{singular} Lagrangians. In
section \ref{sect2} we derive the equivalent mini-superspace model.
In section \ref{sect3} we examine every case for which the
corresponding scaled mini-superspace metric is maximally symmetric
and acquire the analytical solutions for these models. in section
\ref{sect4} we examine some of the properties of the Hubble
parameters that are obtained through the previous mentioned
solutions. Finally in the discussion we sum up our results together
with some concluding remarks.

\section{Preliminary remarks} \label{sect1}
A general relativistic action is of the form \be S=\int\!\!d^4
x\sqrt{-g}\mathcal{L} \ee with $\mathcal{L}$ being the Lagrangian
density of the system and $\sqrt{-g}\, d^4 x$ the infitestimal
space-time volume element. Variation with respect to the space-time
metric $g_{\mu\nu}$ leads to the equations of motion for the
gravitational field \be E_{\mu\nu}=T_{\mu\nu} \ee with $E_{\mu\nu}$
and $T_{\mu\nu}$ the respective contributions from pure geometry
(assuming minimal coupling) and $T_{\mu\nu}$ matter and/or
cosmological constant (if present).

The imposition of certain symmetries in cosmology can lead to a
simplification of the above problem and its reduction to a
mechanical one. Whenever this is the case one is led to a reduced
action \be \label{actred} S_{red}=\int\!\!dt L, \ee where $t$ is the
dynamical variable of the system and $L$ a \emph{singular}
Lagrangian of the general form \be \label{genlag} L = \frac{1}{2N}
G_{\alpha\beta}(q) \dot{q}^\alpha\dot{q}^\beta - N V(q), \ee with
$G_{\alpha\beta}$ ($\det G_{\alpha\beta}\neq 0$) being the
mini-superspace metric, $N$ the lapse function of the base manifold,
$q^\alpha$'s the configuration space variables and
$\dot{q}^\alpha:=\frac{d q^\alpha}{dt}$ their velocities with
respect to the dynamical variable $t$. By adopting the usual
definition of the momenta $p_\alpha:=\frac{\partial L}{\partial
\dot{q}^\alpha}$, $p_N:=\frac{\partial L}{\partial \dot{N}}$ and
following Dirac's algorithm (\cite{Dirac}, \cite{Dirac2},
\cite{Dirac3}) one is led to the Hamiltonian \be H_C =N H + u^N p_N
= N \left(\frac{1}{2} G^{\alpha\beta} p_\alpha p_\beta +V\right) +
u^N p_N \ee where $u^N$ is an arbitrary function. According to the
theory, there exist two first class constraints \be p_N \approx 0,
\quad H\approx 0 \ee that define the restricted phase space.

As shown in \cite{tchris1} and even more generally in
\cite{tchris2}, for singular systems described by Lagrangians of the
form \eqref{genlag}, every conformal vector field of the
mini-supermetric can be used to define integrals of motion linear in
the momenta. In brief, let $\xi$ be a vector field over the
configuration space spanned by the $q$'s, then one can define in
phase space the quantity \be Q=\xi^\alpha p_\alpha, \ee where
$\xi^\alpha$ are the components of $\xi$. If now, \be \label{cond1}
\pound_\xi G_{\alpha\beta}= \omega(q) G_{\alpha\beta} \ee holds, we
distinguish two possibilities:
\begin{itemize}
\item Apart from \eqref{cond1}, the relation
\be \label{cond2}
\pound_\xi V= -\omega(q) V
\ee
also holds, then $Q$ is itself an autonomous integral of motion, because
\be
\frac{dQ}{dt}=\{Q,H_C\} = \omega\, N\, H \approx 0.
\ee
\item The Lie derivative of the potential with respect to $\xi$ has a different conformal factor, i.e.
\be \label{cond2} \pound_\xi V= -\sigma(q) V \ee with $\sigma\neq
-\omega$. Then there can be defined a rheonomic integral of motion
\be I= Q + \int\!\! N \left(\omega(q(t))+\sigma(q(t))\right) V dt
\ee with $q(t)$ being the trajectories obtained by solving the Euler
- Lagrange equations. It is easy to check, that \be \frac{dI}{dt}=
\{Q,H_C\} + N (\omega+\sigma) V =\frac{N \omega}{2}
G^{\alpha\beta}p_\alpha p_\beta + N \omega V = \omega\, N\, H
\approx 0. \ee However, one can argue, that these rheonomic
integrals are not useful, for the purpose of integrating the
equations of motion, since their solutions $q(t)$ need to be known a
priori. Later on we will see that it is possible to overcome this
difficulty, for one such quantity.
\end{itemize}

We can exploit the reparametrization invariance that is exhibited by
theories with Lagrangians of the form of \eqref{genlag} to make the
previous results even clearer. A scaling transformation of the lapse
function $N =\frac{n}{V}$ leads to an equivalent Lagrangian \be L =
\frac{1}{2n} \bbar{G}_{\alpha\beta}(q) \dot{q}^\alpha\dot{q}^\beta -
n \ee and consequently to a Hamiltonian \be H_C = n H + u^n p_n = n
\left(\frac{1}{2} \bbar{G}^{\alpha\beta} p_\alpha p_\beta +1\right)+
u^n p_n \ee with $\bbar{G}_{\alpha\beta}= V G_{\alpha\beta}$. The
autonomous integrals of the system, $Q=\xi^\alpha p_\alpha$, are now
constructed by Killing fields of the scaled supermetric ($\pound_\xi
\bbar{G}_{\alpha\beta}=0$) and their evolution is strictly zero, not
just weakly, i.e. $\{Q,H_C\}=0$. The rheonomic integrals of motion
assume the form \be \label{rheI} I= Q + \int\!\! n\, \omega(q) dt
\ee with $\omega\neq 0$ defined by $\pound_\xi
\bbar{G}_{\alpha\beta}= \omega \bbar{G}_{\alpha\beta}$. It is now
clear from the above relation that, one of these rheonomic integrals
can be utilized prior to the explicit knowledge of $q(t)$, in two
ways: (a) by adoption of a particular gauge choice $n \propto
\frac{1}{\omega(q)}$ or (b) if $\xi$ is a homothecy of the scaled
supermetric $\bbar{G}_{\alpha\beta}$, i.e. $\omega$=const.

It is obvious that, in our analysis, not fixing the time gauge by
selecting a particular lapse function (prior to the derivation of
the equations of motion) is imperative. In the appendix of
\cite{tchris1} we have proved that the gauge fixing of the lapse in
Lagrangian \eqref{genlag} may lead to a loss of some existing
symmetries ($\xi$'s related to integrals of motion). Since
Lagrangians of the form \be \label{reglag} L= \frac{1}{2}
G_{\alpha\beta} \dot{q}^\alpha \dot{q}^\beta - V, \ee that
correspond to the gauge $N =1$, are commonly used in the literature
of mini-superspace cosmology instead of \eqref{genlag}, let us
explain briefly why this is the case: Every physical theory, with
finite degrees of freedom, and its properties are described by an
action principle similar to \eqref{actred}. Lagrangians
\eqref{genlag} and \eqref{reglag} correspond to different physical
theories. If the dimension of $G_{\alpha\beta}$ is $d$, then
\eqref{genlag} corresponds to a (singular) system of $d+1$ degrees
of freedom, while \eqref{reglag} refers to a (regular) system with
$d$ degrees of freedom. The first is invariant under transformations
of the form $t\rightarrow v=h^{-1}(t)$; these transformations are
the remnant of the general coordinate covariance of the base
manifold metric components, i.e. the covariance under an arbitrary
change in the time variable $N(t)^2 dt^2 \rightarrow N(u)^2
\left(\frac{dh}{dv}\right)^2 dv^2 = \widetilde{N}^2(v) dv^2$. The
second system is just a Newtonian-like system i.e. its action is
invariant under only rigid time translations, $t\rightarrow u =
t+\varepsilon$, where $\varepsilon$ is constant. This happens
because, in the latter case, there is no constraint equation of
motion $\frac{\partial L}{\partial N}=0$. As a consequence, the
resulting Hamiltonian is constant but \emph{not necessarily} equal
to zero destroying the reparameterization invariance of the theory,
which in itself is imperative for many of the existing symmetries.
Moreover, in order for someone to acquire the correct solution
space, the condition of the Hamiltonian being zero must be demanded
as an ad hoc condition (not already incorporated in the action
principle).

For all the above mentioned reasons, we \emph{never} gauge fix the
lapse at the Lagrangian level, prior to the derivation of the
symmetries and the equations of motion. Any gauge fixing, where
necessary - for the sake of simplifying the equations - is imposed
strictly after the derivation of symmetries and never inside the
Lagrangian function.

\section{FLRW f(R) cosmology with perfect fluid} \label{sect2}
We assume a Lagrangian density $\mathcal{L}$ consisting of the
gravitational part $\mathcal{L}_g = f(R)$ plus a possible
contribution of matter $\mathcal{L}_m$ \be \label{act1} S= \int\!\!
d^4 x\sqrt{-g} (f(R)+2 \mathcal{L}_m). \ee Variation with respect to
the base manifold metric $g_{\mu\nu}$ leads to the known set of
field equations for the metric $f(R)$ gravity \be \label{feq} f'(R)
R_{\mu\nu}-\frac{1}{2}f(R) g_{\mu\nu}+g_{\mu\nu}
f'(R)^{;\lambda}_{\phantom{;\lambda};\lambda}-f'(R)_{;\mu ;\nu}
=T_{\mu\nu} \ee where $'=\frac{d}{dR}$, ``$;$" stands for the
covariant derivative with respect to $g_{\mu\nu}$ and
$T_{\mu\nu}:=-\frac{2}{\sqrt{g}}\frac{\delta (\sqrt{g} L_m)}{\delta
g^{\mu\nu}}$ is the energy momentum tensor, which is bound to
satisfy \be \label{encon} T^{\mu\nu}_{\phantom{\mu\nu};\mu}=0. \ee

By considering an FLRW space-time \be \label{sptmetric} ds^2 = -
N(t)^2 dt^2 + \frac{a(t)^2}{1-k r^2}dr^2 + a(t)^2 (r^2 d\theta^2+
r^2 \sin\theta d\phi^2) \ee and when one integrates out the
redundant degrees of freedom ($r$, $\theta$, $\phi$), the
contribution of $\sqrt{-g}f(R)$ in \eqref{act1} is just $N a^3
f(R)$. The scalar curvature is \be \label{curv1} R = \frac{6
\left(-a \dot{a} \dot{N}+N \left(a \ddot{a}+\dot{a}^2\right)+k
N^3\right)}{a^2 N^3}, \ee where the dot denotes differentiation with
respect to the time coordinate $t$. In order for \eqref{curv1} to be
reproduced as an equation of motion in the corresponding
minisuperspace model, we add it as a constraint in the ensuing
Lagrangian and then properly fix the Lagrange multiplier
(\cite{Vilenkin}, \cite{Huang}). In short, we assume a Lagrangian
for the gravitational part that is of the form \be \label{Lagg1}
L_{g_1} = N a^3 f(R) + \lambda(t) \left(R -\frac{6 \left(-a \dot{a}
\dot{N}+N \left(a \ddot{a}+\dot{a}^2\right)+k N^3\right)}{a^2
N^3}\right). \ee The equation of motion with respect to $R$ is just
\be \frac{\partial L_{g_1}}{\partial R} =0 \Rightarrow \lambda(t) =
N a^3 f'(R). \ee If we insert this result into \eqref{Lagg1} and
eliminate the acceleration term $\ddot{a}$ by subtracting a total
derivative of time, we are led to the following Lagrangian for the
gravitational part \be L_g = -\frac{6}{N} \left(a f'(R) \dot{a}^2+
a^2 f''(R) \dot{a}\dot{R}\right) + N a \left(a^2 \left(f(R)-R
f'(R)\right)+6 k f'(R)\right). \ee

For the matter part we consider a perfect fluid described by a
barotropic equation of state $p(t) = \gamma \rho(t)$, where $p$ is
the pressure, $\rho$ the energy density and $\gamma$ a constant. The
energy momentum tensor is \be T_{\mu\nu} = (\rho+p) u_\mu u_\nu + p
g_{\mu\nu}, \ee where $u^\mu$ is the 4-velocity of the fluid with
$u^\mu u_\mu=-1$. The equation derived by the conservation of energy
\eqref{encon} is \be 3 (1+\gamma) \rho \dot{a}+a \dot{\rho} =0 \ee
with the well known solution \be \label{ensol} \rho = m
a^{-3(1+\gamma)}, \ee where $m$ is the constant of integration. The
Lagrangian density for the perfect fluid is consisted solely of the
energy density $\rho$, i.e. $\mathcal{L}_m= -\rho$. By considering
\eqref{ensol} we make the assumption that the matter contribution in
the Lagrangian for the equivalent mechanical system is \be L_m = -m
N a^{-3\gamma}. \ee So the full Lagrangian under consideration
becomes \be \label{Lag}
\begin{split}
L = L_g+2 L_m = & -\frac{6}{N} \left(a f'(R) \dot{a}^2+ a^2 f''(R) \dot{a}\dot{R}\right)  \\
& +N a \left(a^2 \left(f(R)-R f'(R)\right)+6 k f'(R)- 2 m a^{-(1+3\gamma)}\right).
\end{split}
\ee
It is an easy task to check, that the Euler - Lagrange equations of \eqref{Lag}
\begin{subequations}
\begin{align} \label{eulN}
\frac{\partial L}{\partial N} & =0 \\ \label{eula}
\frac{d}{dt}\left(\frac{\partial L}{\partial \dot{a}}\right) -\frac{\partial L}{\partial a} & =0 \\ \label{eulR}
\frac{d}{dt}\left(\frac{\partial L}{\partial \dot{R}}\right) -\frac{\partial L}{\partial R} & =0
\end{align}
\end{subequations}
are equivalent to the field equations \eqref{feq}, with \eqref{eulR}
revealing as its solution equation \eqref{curv1}, which is given by
the definition of $R$. As a result, we state that Lagrangian
\eqref{Lag} is valid, hence we can proceed with the mini-superspace
analysis.

At this point, we choose to exploit the reparametrization invariance
exhibited by Lagrangian \eqref{Lag} through a re-scaling of the
lapse function ($N\rightarrow n$) as follows \be \label{scaleN} N =
-\frac{n\, a^{3 \gamma}}{6 k a^{3 \gamma +1} f'(R)-R a^{3(\gamma
+1)} f'(R)+a^{3( \gamma +1)} f(R)-2 m}. \ee We have already stated
that this corresponds to a valid transformation ($t\rightarrow
h(t)$) for a relativistic theory. Under the above transformation the
Lagrangian assumes the form \be L = \frac{A(a,R)}{n} \dot{a}^2  +
\frac{B(a,R)}{n} \dot{a} \dot{R} - n \ee with
\begin{subequations}
\begin{align}
A(a,R) & = 6\, a f'(R) \left(6 k a f'(R)+a^3 \left(f(R)-R f'(R)\right)-2 m a^{-3 \gamma }\right) \\
B(a,R) & = 6\, a^{2-3 \gamma } f''(R) \left(6 k a^{3 \gamma +1} f'(R)+a^{3 (\gamma +1)} \left(f(R)-R f'(R)\right)-2 m\right).
\end{align}
\end{subequations}
As already mentioned in the previous section, it is in this
parametrization of the constant potential, that the Killing fields
of the scaled mini-supermetric \be \label{supmet} G_{\alpha\beta} =
\begin{pmatrix}
2 A & B \\ \\
B & 0
\end{pmatrix}
\ee generate autonomous integrals of motion in phase space (for
simplicity we avoid the bar symbolism that we used in the previous
section for this mini-supermetric).

The scaled Hamiltonian reads
\be
H_C= \frac{n}{2}G^{\alpha\beta}p_\alpha p_\beta +n + u^n p_n
\ee
where $\alpha,\beta={a,R}$ with
\be \label{momenta}
p_a = \frac{\partial L}{\partial \dot{a}}, \quad p_R = \frac{\partial L}{\partial \dot{R}} \quad.
\ee
The constraint space is defined by the two first class constraints
\begin{subequations}
\begin{align}
p_n & \approx 0 \\
H  = \frac{1}{2}G^{\alpha\beta}p_\alpha p_\beta +1 & \approx 0
\end{align}
\end{subequations}
and each Killing vector field of \eqref{supmet}, $\pound_{\xi_J}
G_{\alpha\beta}=0$, defines the autonomous integral of motion
$Q_J=\xi_J^\alpha\, p_\alpha$, because $\{Q_J,H_C\} = 0$; while a
homothecy of $G_{\alpha\beta}$ ($\pound_{\xi_h}
G_{\alpha\beta}=G_{\alpha\beta}$) leads to the rheonomic integral
$I_h=Q_h+\int\!\! n dt = \xi_h^\alpha \, p_\alpha+\int\!\! n dt$.
Thus in each case we can define the following set of equations \be
\label{consteqs} Q_J = \kappa_J \quad \text{and} \quad I_h =
\kappa_h, \ee where $\kappa_J$, $\kappa_h$ are constants with $J$
counting the number of possibly existing Killing fields.

\section{Specific models} \label{sect3}
The number of Killing fields of \eqref{supmet} is upper bounded by
its dimension. As it is known for a metric of dimension $d$, the
maximum number of possible isometries is $\frac{d (d+1)}{2}$. Thus,
in our case, there can be at most three Killing fields of
$G_{\mu\nu}$ and, accordingly, three autonomous integrals of motion
linear in the momenta. Their number depends on the choice of $f(R)$,
$k$ and $\gamma$ since they are the only free parameters inside the
scaled mini-supermetric. In other words, it is their choice that
determines the geometry of mini-superspace.

We choose to focus our analysis on those models that correspond to a
maximally symmetric superspace, i.e. $G_{\alpha\beta}$ admits the
maximum number of Killing vectors. Of course, this does not
constitute a physical argument, it is a rather cynical choice so as
to have in our disposal as many first integrals there is possible,
in order to integrate the equations of motion. Our mini-superspace
is two dimensional, hence, for maximal symmetry, the space needs to
be either flat or of constant non zero scalar curvature
$\mathcal{R}$.

One can easily check, that the scalar curvature corresponding to
$G_{\alpha\beta}$ becomes a non zero constant only for a linear
function $f(R)=c_1 R + c_2$ and $k=0$ (driving also $\gamma$ to
particular values). Hence, restricting ourselves to all other cases
of $f(R)$ we need to investigate those models with a flat
mini-supersace. As a starting point we distinguish two major cases,
$k=0$ and $k\neq 0$.

\subsection{Spatially flat models ($k=0$)}
By rejecting the possibility for $f(R)$ being a linear function of
$R$, the mini-superspace is flat, i.e.
$\mathcal{R}_{\alpha\beta\rho\tau}=0$ iff \be \label{flatcond}
a^{3(\gamma+1)}(f(R) \left(R  f''(R)+f'(R)\right)-R  f'(R)^2)-2 m
\left(f'(R)-8 R  f''(R)\right) =0. \ee Now, we have to proceed to a
further separation of possible cases, according to what happens with
$a^{3(\gamma+1)}$ and its coefficient.

\subsubsection{Case $\gamma=-1$}
In this case the matter contribution in the Lagrangian corresponds
to that of a cosmological constant, since $L_m=-Na^3 m =
-\sqrt{-g}\, m$, with $m$ being this constant. Under this selection
for $\gamma$, equation \eqref{flatcond} holds for \be f(R) = 2m +
\lambda R^\mu, \ee with $\lambda$, $\mu$ constants ($\mu\neq 0, 1$).
This choice results in the elimination of the cosmological constant
$m$ in the full Lagrangian \eqref{Lag}. Henceforth, the
corresponding model becomes equivalent to a vacuum model with
$f(R)=\lambda R^\mu$. Thus, we can state that, a cosmological
constant cannot produce (at least not with an FLRW line element) a
maximally symmetric mini-superspace. If one uses the latter as a
requirement, it results in a necessary annihilation of $m$ from the
action itself. For reasons that will become evident later on in the
analysis we have to distinguish several subcases.
\begin{itemize}
\item $\mu\neq \frac{1}{2}$ and $\mu\neq \frac{5}{4}$
\end{itemize}
The three Killing vectors of the scaled mini-supermetric are
\begin{equation} \label{Kill1}
\xi_1  = a \partial_a +\frac{6R}{1-2\mu} \partial_R, \quad
\xi_2  = a^{\frac{4-3\mu}{\mu-1}} \partial_a - \frac{a^{\frac{5-4\mu}{\mu-1}}R}{\mu-1} \partial_R , \quad
\xi_3  = a^{\frac{1-2\mu}{\mu-1}} R^{2(1-\mu)} \partial_R,
\end{equation}
where from the form of $\xi_1$ it becomes evident why we excluded
the value $\mu=\frac{1}{2}$. Additionally, there exists a homothetic
vector ($\pound_{\xi_h}G_{\alpha\beta}=G_{\alpha\beta}$) \be
\label{hom1} \xi_h = \frac{a}{6} \partial_a. \ee Thus, we can define
three autonomous integrals of motion linear in the momenta \be
\label{autintdef} Q_J = \xi_J^\alpha \,  p_\alpha, \quad J=1,2,3,
\quad \alpha=a,R \ee and a rheonomic integral due to the homothetic
vector field \be \label{rehintdef} I_h =Q_h+\int\!\! n(t) dt
=\xi_{h}^\alpha \,  p_\alpha + \int\!\! n(t) dt, \quad \alpha=a,R.
\ee It can easily be verified that $\dot{Q}_J = \{Q_J,H_C\}=0$ for
$J=1,2,3$ and $\dot{I}_h= \{Q_h,H_C\}+n(t)\approx 0$, thus on the
solution space, relations
\begin{subequations}
\begin{align} \label{autint}
Q_J &= \kappa_J \\ \label{rehint}
Q_h+\int\!\! n(t) dt &= \kappa_h
\end{align}
\end{subequations}
with $\kappa_J$, $\kappa_h$ being constants hold. By substituting in
\eqref{autint} and \eqref{rehint} the momenta with respect to the
configuration space variables and the velocities from
\eqref{momenta}, we are led to the following set of equations:
\begin{subequations} \label{eqset1}
\begin{align}  \label{eqset11}
\kappa_1 (1-2 \mu) n R^2-6 \lambda^2 (\mu-1) \mu a^5R^{2\mu} \left(\left(2 \mu^2-3 \mu+1\right) a \dot{R}-2 (\mu-2) R \dot{a}\right) &=0 \\  \label{eqset12}
6 \lambda^2 (\mu-1) \mu a^{\frac{\mu}{\mu-1}} R^{2 \mu} \left(R \dot{a}+(\mu-1) a \dot{R}\right)+\kappa_2 n R^2 &=0 \\  \label{eqset13}
6 \lambda^2 (\mu-1)^2 \mu a^{\frac{3\mu-4}{\mu-1}} \dot{a}+\kappa_3 n & =0 \\  \label{eqset14}
n R^2 \left(\int\!\!ndt-\kappa_h\right)-\lambda^2 (\mu-1) \mu a^5 R^{2 \mu} \left(2 R \dot{a}+(\mu-1) a \dot{R}\right) &=0.
\end{align}
\end{subequations}
By solving equation \eqref{eqset13} with respect to $n$ we get \be
\label{nrel1} n = -\frac{6 \lambda^2 (\mu-1)^2 \mu
a^{\frac{3\mu-4}{\mu-1}} \dot{a}}{\kappa_3}. \ee Two remarks are in
order: (a) $\kappa_3$ cannot be zero, for then $a(t)$ would be
constant, $R(t)$ zero and the resulting space-time would correspond
to a trivial solution with $\gamma=-1$ and $f(R)=2\, m$, (b) the
fact that we solved \eqref{eqset13} with respect to $n$ does not
constitute a gauge choice. This is because relations \eqref{eqset1}
are valid in any time gauge (as long as the field equations hold),
no matter what is the particular functional form of $n$, $a$ or $R$.

By using \eqref{nrel1} in the remaining equations, the gauge freedom
passes over to one of the last two degrees of freedom, $a$ and $R$.
So, we can choose either of them as an explicit function of time. A
rather convenient choice is \be \label{scf1} a(t) = t R^{1-\mu}, \ee
for then, equation \eqref{eqset11} can be integrated to give \be
\label{rho1} R(t) = \frac{e^{\sigma} t^{\frac{4}{4 \mu
-5}}}{\left(\kappa_1 (2 \mu -1)(\mu-1) t^{\frac{1}{\mu -1}}+\kappa_3
(4 \mu -5) t^{\frac{2 \mu }{\mu -1}}\right)^{\frac{1}{4 \mu -5}}}
\ee where $\sigma$ is an integration constant. From $\eqref{rho1}$
we can see why we also had to exclude the case $\mu=\frac{5}{4}$.
Solutions \eqref{nrel1}, \eqref{scf1} and \eqref{rho1} together with
\eqref{scaleN} satisfy the Euler - Lagrange equations \eqref{eula}
and \eqref{eulR} for this model, while the quadratic constraint
equation \eqref{eulN} yields a relation between constants \be
\label{const1} \kappa_3 = 6 \lambda^2 \mu (\mu -1)^3  (2 \mu -1)
e^{(5-4 \mu ) \sigma}. \ee Relations \eqref{nrel1}, \eqref{scf1},
\eqref{rho1} and \eqref{const1} constitute the solution of the given
system. If we substitute them into \eqref{eqset12} and
\eqref{eqset14} we can evaluate the value of two more constants,
$\kappa_2=\frac{e^{(4 \mu -5) \sigma}}{2 \mu ^2-3 \mu +1}$ and
$\kappa_h = \frac{\kappa_1(1-2 \mu) }{6 (4 \mu -5)}$. The ensuing
line element is \eqref{sptmetric} ($k=0$) with a lapse function \be
\label{sollap1} N = 6 \lambda\, \mu \, (\mu -1)  (2 \mu -1) e^{-2
(\mu -1) \sigma } t^{\frac{2 \mu -3}{(\mu -1) (4 \mu -5)}}
\left(\kappa_1 (2\mu-1)(\mu-1) t^{\frac{1}{\mu -1}}+\kappa_3 (4 \mu
-5) t^{\frac{2 \mu }{\mu -1}}\right)^{\frac{3-2 \mu }{4 \mu -5}} \ee
and a scale factor given by \eqref{scf1} with the substitution of
\eqref{rho1}.

\begin{itemize}
\item $\mu= \frac{1}{2}$
\end{itemize}
This time the three Killing vectors of the supermetric are
\begin{equation} \label{Kill2}
\xi_1  = a \partial_a +6 R\ln\left(\frac{a^2}{R}\right)  \partial_R, \quad
\xi_2  = \frac{1}{a^5} \partial_a -  \frac{2 R}{a^6} \partial_R , \quad
\xi_3  = R \partial_R.
\end{equation}
The latter two are the same as the $\xi_2$ and $\xi_3$ from
\eqref{Kill1} when one sets $\mu=\frac{1}{2}$, but the first one
cannot be obtained this way. Apart from these, there also exists a
homothetic vector which is identical to \eqref{hom1}. Thus, it is
possible to define in the same way the linear integrals of motion
\eqref{autintdef}, \eqref{rehintdef} and consequently the respective
four equations \eqref{autint} and \eqref{rehint} that yield
following the system:
\begin{subequations} \label{eqset2}
\begin{align}  \label{eqset21}
3 \lambda^2 a^5 \left(2 R \dot{a} \left(3 \log\left(\frac{a^2}{R}\right)-2\right)+a \dot{R}\right)+4 \kappa_1 n  R &=0 \\  \label{eqset22}
\frac{3 \lambda^2 \dot{a}}{2 a n}-\kappa_2-\frac{3 \lambda^2 \dot{R}}{4 n R} &=0 \\  \label{eqset23}
-\frac{3 \lambda^2 a^5 \dot{a}}{4 n}-\kappa_3 & =0 \\  \label{eqset24}
\int\!\! n dt + \frac{\lambda^2 a^5 \left(4 R \dot{a}-a \dot{R}\right)-8 \kappa_h n R}{8 n R} &=0.
\end{align}
\end{subequations}

Here the situation is quite simpler than before, because the system
can be solved algebraically for two of the three degrees of freedom
and their derivatives, leaving unuttered the gauge freedom (for
additional examples see \cite{tchris2} and \cite{tchris3}). We
choose to solve system \eqref{eqset2} with respect to $\int\!\!n
dr$, $n$, $R$ and $\dot{R}$. As a result we get
\begin{subequations} \label{solset1}
\begin{align}  \label{intn1}
\int\!\! n dt &= -\frac{1}{6} \kappa_2 a^6+ \kappa_h +\frac{\kappa_3}{3} \\  \label{soln1}
n &= -\frac{3 \lambda^2 a^5 \dot{a}}{4 \kappa_3} \\  \label{rho2}
R & = a^2 e^{-\frac{-\kappa_2 a^6+ \kappa_1+ 2 \kappa_3}{6 \kappa_3}} \\  \label{dotrho2}
\dot{R} &=\frac{a  \dot{a} \left(\kappa_2 a^6+2 \kappa_3\right) e^{-\frac{-\kappa_2 a^6+\kappa_1+2 \kappa_3}{6 \kappa_3}}}{\kappa_3}.
\end{align}
\end{subequations}
The consistency condition $\dot{R} = \frac{d R}{dt}$ is identically
satisfied, while $n= \frac{d}{dt}\int\!\! n dt$ leads to a relation
between constants \be \label{const2} \kappa_2 = \frac{3 \lambda^2}{4
\kappa_3}. \ee This value of $\kappa_2$ is a realization of the
constraint equation $H\approx 0$ (see \cite{tchris2} and
\cite{tchris3}).

Expressions \eqref{soln1}, \eqref{rho2} and \eqref{const2} (with the
use of \eqref{scaleN}) are solutions of the Euler - Lagrange
equations $\eqref{eulN}$, $\eqref{eula}$ and $\eqref{eulR}$ for this
model. Henceforth, they also solve the field equations \eqref{feq}
under the specific requirements we have made. The corresponding line
element is \be \label{solvac2} ds^2 = -\left(\frac{3 \lambda a(t)
\dot{a}(t) e^{\frac{4 \kappa_3 (\kappa_1+2 \kappa_3)-3 \lambda^2
a(t)^6}{48 \kappa_3^2}}}{2 \kappa_3}\right)^2 dt^2 +  a(t)^2 (dr^2
+r^2 d\theta^2+ r^2 \sin\theta d\phi^2) \ee with $a(t)$ being an
arbitrary function, since we did not need to make use of the gauge
freedom for the integration of system \eqref{eqset2}.

\begin{itemize}
\item $\mu= \frac{5}{4}$
\end{itemize}
This value of the exponent leads to the following three Killing vectors in the configuration space:
\begin{equation} \label{Kill3}
\xi_1  = a \partial_a -4 R  \partial_R, \quad
\xi_2  = a \ln (a) \partial_a - \frac{2}{3} R (1+ 6 \ln (a)) \partial_R , \quad
\xi_3  = \frac{1}{a^6 \sqrt{R}} \partial_R,
\end{equation}
while the homothetic vector remains the same as in the previous cases.

In the usual way we produce the ensuing set of equations \eqref{consteqs} from the definition of the first integrals
\begin{subequations} \label{eqset3}
\begin{align}  \label{eqset31}
15 \lambda^2 a^5 \sqrt{R} \left(4 R \dot{a}+a \dot{R}\right)+32 \kappa_1 n &=0 \\  \label{eqset32}
5 \lambda^2 a^5 \sqrt{R} \left(2 R \dot{a} (6 \ln (a)-1)+3 a \ln (a) \dot{R}\right)+32 \kappa_2 n &=0 \\  \label{eqset33}
-\frac{15 \lambda^2 \dot{a}}{32 a n}-\kappa_3 & =0 \\  \label{eqset34}
5 \lambda^2 a^5 \sqrt{R} \left(8 R \dot{a}+a \dot{R}\right)+64 n \left(\kappa_h-\int\!\! n dt\right) &=0.
\end{align}
\end{subequations}
Again the system can be solved algebraically with respect to $\int\!\!n dr$, $n$, $R$ and $\dot{R}$ and doing so we get
\begin{subequations} \label{solset2}
\begin{align}  \label{intn2}
\int\!\! n dt &= -\kappa_1 \ln (a)+ \kappa_h - \frac{\kappa_1}{6}+\kappa_2 \\  \label{soln2}
n &= -\frac{15 \lambda^2 \dot{a}}{32 \kappa_3 a} \\  \label{rho3}
R & = \frac{\left(\frac{3}{2}\right)^{2/3} (\kappa_1 \ln (a)- \kappa_2)^{2/3}}{\kappa_3^{2/3} a^4} \\  \label{dotrho3}
\dot{R} &= \frac{\left(\frac{2}{3}\right)^{1/3} \dot{a} (-6 \kappa_1 \ln (a)+ \kappa_1 +6 \kappa_2)}{\kappa_3^{2/3} a^5 \left(\kappa_1 \ln (a)- \kappa_2\right)^{1/3}}.
\end{align}
\end{subequations}
Once more, the consistency condition $\dot{R}= \frac{d R}{dt}$ is
identically satisfied, while $n = \frac{d}{dt}\int\!\! n dt$ leads
to the relation \be \label{const3} \kappa_3 = \frac{15 \lambda^2}{32
\kappa_1}. \ee As is expected, \eqref{soln2}, \eqref{rho3} and
\eqref{const3} together with \eqref{scaleN} solve the field
equations for this particular model. The resulting line element is
\be \label{solvac3} ds^2 = -\left(\frac{5^{5/6} \kappa_1^{1/6}
\lambda^{2/3} a(t) \dot{a}(t)}{2^{4/3} (\kappa_1 \ln
\left(a(t)\right)-\kappa_2)^{5/6}} \right)^2 dt^2  + a(t)^2 (dr^2 +
r^2 d\theta^2+ r^2 \sin\theta d\phi^2) \ee with $a(t)$ remaining an
arbitrary function of time $t$.

All the solutions obtained in this section (\eqref{solvac2} for
$\mu=1/2$, \eqref{solvac3} for $\mu=5/4$ and the set \eqref{scf1},
\eqref{rho1} and \eqref{sollap1} for all the other values of $\mu$),
that actually correspond to the vacuum case with $f(R)=\lambda
R^\mu$, have also been obtained by different means in
\cite{Clifton}.

\subsubsection{Case $\gamma \neq -1$, $\gamma \neq \pm \frac{2}{3}$ and $\gamma \neq - \frac{4}{3}$}
The reason why we exclude the values $\pm \frac{2}{3}$ and
$-\frac{4}{3}$ for $\gamma$ will become evident later on in the
analysis. In order for \eqref{flatcond} to be zero, one has to
eliminate the coefficients of different powers of $a$. When $\gamma
\neq -1$, at first we set \be \label{fRexp} f(R) = \lambda R^\mu,
\ee for the cancelation of the coefficient of $a^{3(\gamma+1)}$.
Then the requirement for the full vanishing of \eqref{flatcond}
leads to a relation between $\mu$ and $\gamma$, i.e. to a
correspondence between gravity and equation of state, \be
\label{muval} \mu = \frac{3 (\gamma +1)}{3 \gamma +2}. \ee As we can
see the value $\gamma = -\frac{2}{3}$ would lead to a non flat
mini-superspace which would not be maximally symmetric and thus out
of our scope.

Under the assumptions \eqref{fRexp} and \eqref{muval} the mini-supermetric exhibits three Killing vectors
\begin{subequations} \label{Kill4}
\begin{align}
\xi_1  &= a \partial_a -\left(\frac{6 (3 \gamma +2) \left((\gamma -1) (3 \gamma +2) (3 \gamma +4) m R -\lambda  a^{3 (\gamma +1)} R^{\frac{5+6 \gamma}{2+3\gamma}}\right)}{(3 \gamma +4) \left(\lambda  a^{3 (\gamma +1)} R^{\frac{3(\gamma+1)}{2+3\gamma}}+(6 \gamma +4) m\right)}\right)  \partial_R \\
\xi_2  &= a^{3\gamma-1} \partial_a - (3 \gamma +2) R  a^{3 \gamma -2} \partial_R  \\
\xi_3  &= \left(\frac{5 (3 \gamma +2) R^{\frac{1+3\gamma}{2 +3\gamma}}}{a (3 \gamma +4) \left(\lambda  a^{3 (\gamma +1)} R^{\frac{3(\gamma+1)}{2+3\gamma}}+(6 \gamma +4) m\right)}\right) \partial_R,
\end{align}
\end{subequations}
and a homothetic Killing vector \be \label{hom2} \xi_h =
\frac{a}{2-3\gamma} \partial_a - \frac{2+3\gamma}{2-3\gamma} R
\partial_R. \ee From the form of \eqref{Kill4} and \eqref{hom2} it
becomes clear that cases $\gamma=\frac{2}{3}$ and $\gamma =
-\frac{4}{3}$ should be treated separately as special cases. For
now, we proceed by excluding these particular values.

By the use of \eqref{autintdef}, \eqref{rehintdef} and the
definition of the momenta, we can obtain four equations involving
$a$, $\dot{a}$, $R$, $\dot{R}$, $\int\!\! n dt$ and $n$.
Unfortunately, due to the arbitrariness of $\gamma$, it is not
possible to solve those equations algebraically (i.e. without making
a gauge choice). The simplest equation is the one defined by $\xi_3$
which yields \be \frac{90 (\gamma +1) \lambda  a^{1-3 \gamma
}\dot{a}}{(3 \gamma +2)^2 (3 \gamma +4) n}+\kappa_3 =0. \ee If we
solve the latter with respect to $n$ we get \be \label{nrel2} n =
-\frac{90 (\gamma +1) \lambda  a^{1-3 \gamma } \dot{a}}{(3 \gamma
+2)^2 (3 \gamma +4) \kappa_3}, \ee with $\kappa_3\neq 0$ for the
same reason that we mentioned in the first subsection. As we stated
previously \eqref{nrel2} is not a gauge choice, this freedom is
transported to the degrees of freedom $a$ or $R$. As gauge we choose
\be \label{scf2} a = t R^{-\frac{1}{2+3\gamma}}. \ee By substitution
of  \eqref{nrel2} and \eqref{scf2} in the equation defined by $Q_1 =
\kappa_1$ we get a first order ODE for $R(t)$, which upon
integration yields \be \label{rho4} R =  \frac{\sigma \, t^{3 \gamma
+2}}{\left(5 \kappa_1 - (3 \gamma -2) \kappa_3 t \left(2 \left(9
\gamma ^2+18 \gamma +8\right) m+\lambda  t^{3 \gamma
+3}\right)\right)^{\frac{2+ 3 \gamma}{2 -3 \gamma}}}, \ee with
$\sigma$ being a constant. Relations \eqref{nrel2}, \eqref{scf2},
\eqref{rho4} and by the use of \eqref{scaleN} solve the spatial
equations of motion $\eqref{eula}$ and $\eqref{eulR}$, while the
quadratic constraint \eqref{eulN} gives rise to a relation between
constants \be \label{const4} \kappa_3 = \frac{450 (\gamma +1)
\lambda  \sigma^{\frac{3\gamma-2}{3 \gamma +2}}}{(3 \gamma +2)^2 (3
\gamma +4)}. \ee The other equations for the integrals of motion,
lead also to relations regarding the respective constants. The
equation defined by $Q_2 = \kappa_2$ gives $\kappa_2=
\frac{1}{5}\sigma^{\frac{3\gamma-2}{3\gamma+2}}$, while
$Q_h+\int\!\! n(t) dt = \kappa_h$ leads to $\kappa_h=0$.

From \eqref{scaleN} we can evaluate the lapse function in line
element \eqref{sptmetric} ($k=0$) to be \be \label{genlap} N =
-\frac{90 (\gamma +1) \lambda  \sigma^{-\frac{2}{3 \gamma +2}}
\left(5 \kappa_1-(3 \gamma -2) \kappa_3 t \left(2 \left(9 \gamma
^2+18 \gamma +8\right) m+\lambda  t^{3 \gamma
+3}\right)\right)^{\frac{3 \gamma }{2-3 \gamma }}}{3 \gamma +2} \ee
with $\kappa_3$ given by \eqref{const4} and $\kappa_1$, $\sigma$
remaining arbitrary constants. The scale factor $a(t)$ is given by
\eqref{scf2} with substitution of \eqref{rho4}. This solution was
also attained in \cite{Clifton} with the same exceptions $\gamma
\neq \pm \frac{2}{3}$ and $\gamma \neq - \frac{4}{3}$.

Since we managed to solve the field equations for a general $\gamma$
(apart from four specific values), we can use $\eqref{genlap}$,
\eqref{scf2}, \eqref{rho4} and \eqref{const4} to evaluate solutions
for specific equations of state (of course we must bear in mind that
$\mu$ is also fixed through \eqref{muval}): For a pressureless
matter $\gamma=0 \Rightarrow \mu =\frac{3}{2}$ we have \be N= -45
\frac{\lambda}{\sigma}, \quad a = \left(\frac{5 \kappa_1+32 \kappa_3
m t+2 \kappa_3 \lambda t^4}{\sigma}\right)^{1/2}, \quad \kappa_3 =
\frac{225 \lambda}{6 \sigma}, \ee for radiation
$\gamma=\frac{1}{3}\Rightarrow \mu=\frac{4}{3}$ we get \be N=
-\frac{40 \lambda  \left(5 \kappa_1 +30 \kappa_3 m t+\kappa_3
\lambda  t^5\right)}{\sigma^{2/3}}, \quad a = \frac{5 \kappa_1+30
\kappa_3 m t+\kappa_3 \lambda t^5}{\sigma^{1/3}}, \quad \kappa_3 =
\frac{40 \lambda }{3 \sigma^{1/3}} \ee and finally for stiff matter
$\gamma=1\Rightarrow \mu=\frac{6}{5}$ the result is \be
\begin{split}
N = -\frac{36 \lambda }{\sigma^{2/5} \left(5 \kappa_1- \kappa_3 t
\left(70 m+\lambda  t^6\right)\right)^3}, & \quad a =
\frac{1}{\sigma^{1/5} \left(5 \kappa_1-\kappa_3 t \left(70 m+\lambda
t^6\right)\right)}, \\ \kappa_3 & = \frac{36 \lambda
\sigma^{1/5}}{7}.
\end{split}
\ee

\subsubsection{Case $\gamma=\frac{2}{3}$} \label{new1}
This is a special case of a model with $f(R)$ given by
\eqref{fRexp}. Under this specific choice for $\gamma$, and through
\eqref{muval}, we are led to $\mu=\frac{5}{4}$. The resulting
mini-superspace is of course (since we chose so) flat and its metric
has the following three Killing vectors
\begin{subequations} \label{Kill5}
\begin{align}
\xi_1  &= a \partial_a -4 R  \partial_R \\
\xi_2  &= a \ln(a) \partial_a -\frac{2 R  \left(a^5 \lambda R^{5/4}+6 \ln(a) \left(a^5 \lambda  R^{5/4}+8 m\right)+48 m\right)}{3 \left(a^5 \lambda  R^{5/4}+8 m\right)} \partial_R  \\
\xi_3  &= \frac{10 R^{3/4}}{3 a^6 \lambda  R^{5/4}+24 a m} \partial_R,
\end{align}
\end{subequations}
along with the homothetic vector \be \xi_h = \frac{2 a^5 \lambda
R^{9/4}+96 m R}{3 a^5 \lambda  R^{5/4}+24 m} \partial_R \ee The
equations corresponding to those four quantities that are constants
of motion cannot be given in closed form by algebraically solving
for two of the degrees of freedom and their derivatives. Thus, we
are obligated to proceed by a suitable gauge fixing. The same
procedure is applied, first we solve equation $Q_3= \kappa_3$ with
respect to the scaled lapse $n$ \be \label{nrel3} n = -\frac{25
\lambda \dot{a}}{16 \kappa_3 a}, \ee then we adopt the gauge choice
\be \label{scf3} a = t R^{-1/4} \ee and equation $Q_1 = \kappa_1$
can be integrated to give \be \label{rho5} R = \sigma t^4
e^{-\frac{4 \kappa_3 t \left(48 m+\lambda  t^5\right)}{5 \kappa_1}}.
\ee Relations \eqref{nrel3}, \eqref{scf3} and \eqref{rho5} solve he
spatial field equations of the model under consideration. The
quadratic constraint reveals a relation between constants \be
\label{const5} \kappa_3 = \frac{25 \lambda }{16 \kappa_1} \ee which
completes the solution of the equations of motion. The constant
values of the other integrals can be evaluated to be \be \kappa_2 =
-\kappa_h = -\frac{1}{4} \kappa_1 \ln (\sigma). \ee Finally, the
lapse in line element \eqref{sptmetric} (with $k=0$) can be
calculated through \eqref{scaleN}, the latter yields \be N=
-\frac{15 \lambda e^{\frac{5 \lambda  t \left(48 m+\lambda
t^5\right)}{8 \kappa_1^2}}}{2 \kappa_1 \sqrt{\sigma}} \ee and of
course the scale factor is given by \eqref{scf3}. To the best of our
knowledge this solution has not been previously presented in the
relevant literature.

\subsubsection{Case $\gamma=-\frac{4}{3}$} \label{new2}
One can proceed exactly as in the previous subcase. This time $\mu
=\frac{1}{2}$, the mini-supermetric has the three Killing vectors
\begin{subequations} \label{Kill6}
\begin{align}
\xi_1  &= a \partial_a + \left(\frac{-12 \lambda  R^{3/2} \ln (a)+56 a m R +6 \lambda  R^{3/2} \ln (R)}{4 a m-\lambda  \sqrt{R}}\right)  \partial_R \\
\xi_2  &= a^{-5} \partial_a + \frac{2 R}{a^6} \partial_R  \\
\xi_3  &= \frac{R^{3/2}}{4 a m-\lambda  \sqrt{R}} \partial_R,
\end{align}
\end{subequations}
and the homothetic vector
\be
\xi_h = \frac{a}{6} \partial_a + \frac{R}{3} \partial_R.
\ee
By following the gauge fixing approach we arrive at the solution
\begin{subequations}
\begin{align}
n = & \frac{3 \lambda  a^5 \dot{a}}{4 \kappa_3} \\
a = & t \sqrt{R} \\
R = & \frac{\left(\sigma \kappa_2-48 \kappa_3 m t+12 \kappa_3 \lambda  \ln(t)\right)^{1/3}}{\kappa_2^{1/3} t^2} \\
\kappa_2 = & -\frac{3 \lambda }{4 \kappa_1}.
\end{align}
\end{subequations}
We note that, from the equations defined by the integrals, we chose
to solve $Q_3=\kappa_3$ and $Q_2 =\kappa_2$. The other two equations
give the relations \be \kappa_1 =\kappa_2 \sigma +2 \kappa_3
\lambda, \quad \kappa_h = 0. \ee Equation \eqref{scaleN} yields the
lapse function of the space time line element, which is \be N =
\frac{4 \kappa_3 \lambda ^{2/3}}{\left(\lambda  \sigma +64
\kappa_3^2 m t-16 \kappa_3^2 \lambda  \ln (t)\right)^{2/3}}. \ee
Again the solution here obtained is, to the best of our knowledge,
new.

\subsection{Non vanishing $k$} \label{new3}
For a non vanishing $k$, the mini-superspace is flat if
\be \label{flatcond2}
\begin{split}
6 a^2 f''(R) \left(-f''(R) \left(2 m \left(18 \gamma  k-a^2 (3 \gamma +2) R \right)-a^{3 (\gamma +1)} f(R) \left(a^2 R -18 k\right)\right) \right. \\
\left. +a^{3 (\gamma +1)} \left(-\left(a^2 R -6 k\right)\right) f'(R)^2+a^2 f'(R) \left(a^{3 (\gamma +1)} f(R)-2 m\right)\right) = 0
\end{split}
\ee We ignore the case of $f(R)$ being a linear function of $R$, and
by taking the coefficient of $a^{7+3\gamma}$ we see that again
$f(R)$ must be of the form $f(R) = \lambda R^\mu$. In order to
proceed we consider the coefficient of $a^{5+3\gamma}$ that leads to
the relation \be \mu \left(2 \mu^2-5 \mu +3\right)=0. \ee We again
do not take into account the solutions $\mu=0$ and $\mu=1$, since
they are trivial for the theory, so we are left with just $\mu =
\frac{3}{2}$. Under these assumptions, equation \eqref{flatcond2}
leads finally to $\gamma =0$. We note here, that the specific
choices $\gamma=-\frac{7}{3}$ and $\gamma= - \frac{5}{3}$, which
could change the coefficient arrangements in \eqref{flatcond2}, lead
to trivial choices for $\mu$ and $k$, i.e. $\mu=0,1$ and $k=0$. To
summarize, the requirement that we must have a maximally symmetric
mini-superspace in the $k \neq 0$ case leaves, as the only option, a
pressureless perfect fluid and a (modified) gravitation theory that
is described by \be f(R) = \lambda R^{3/2}. \ee

The three Killing vectors of the mini-supermetric for the model under consideration are
\begin{subequations} \label{Kill7}
\begin{align}
\xi_1  &= \frac{a^2+1}{2 a} \partial_a  -\frac{R  \left(a \lambda  \sqrt{R} \left(a^2 \left(3 a^2+2\right) R -36 \left(2 a^2+1\right) k\right)+8 \left(3 a^2+1\right) m\right)}{2 a^2 \left(a \lambda  \sqrt{R} \left(a^2 R -18 k\right)+4 m\right)}  \partial_R \\
\xi_2  &= \frac{a^2-1}{2 a} \partial_a + \frac{a \lambda  R^{3/2} \left(36 \left(2 a^2-1\right) k+\left(2-3 a^2\right) a^2 R \right)+8 \left(1-3 a^2\right) m R}{2 a^2 \left(a \lambda  \sqrt{R} \left(a^2 R -18 k\right)+4 m\right)} \partial_R  \\
\xi_3  &= \frac{\sqrt{R}}{a \left(a \lambda  \sqrt{R} \left(a^2 R -18 k\right)+4 m\right)} \partial_R.
\end{align}
\end{subequations}
Additionally, there exists a homothetic vector \be \label{hom7}
\xi_h = \frac{a}{2} \partial_a - R \partial_R . \ee The system of
equations that corresponds to the constant integrals of motion that
are constructed by \eqref{Kill7} and \eqref{hom7} cannot be solved
algebraically, thus we follow the procedure that entails a
convenient gauge choice. By solving equation $Q_3 = \kappa_3$ with
respect to $n$ we get \be \label{nrel4} n = -\frac{9 \lambda  a
\dot{a}}{4 \kappa_3} . \ee Substitution of \eqref{nrel4} into $Q_1 =
\kappa_1$ and under the gauge choice \be \label{scf4} a = t
R^{-1/2}, \ee yields a first order ODE for $R(t)$ which can be
integrated to give \be \label{rho6} R = \frac{t^2}{\sigma \left(2
\kappa_1 -36 \kappa_3 k \lambda   t^2+16 \kappa_3 m  t+\kappa_3
\lambda   t^4 \right)-1}, \ee with $\sigma$ being the constant of
integration.

The spatial field equations of this model are satisfied by
\eqref{nrel4} (with the use also of \eqref{scaleN}), \eqref{scf4}
and \eqref{rho6}. The quadratic constraint equation as in the
previous cases leads to a relation between constants \be \kappa_3 =
\frac{9 \lambda  \sigma}{4} \ee and that completes the solution of
the model. Now, it is an easy task to also compute the values of the
other two constants of motion \be \kappa_2 = \frac{\kappa_1
\sigma-1}{\sigma}, \quad \kappa_h = 0. \ee The resulting lapse in
line element \eqref{sptmetric} is (relation \eqref{scaleN}) \be N =
- 9 \lambda \sigma \ee and the scale factor is given by \eqref{scf4}
under the substitution of \eqref{rho6}. This solution is also new -
in its generality - to the best of our knowledge. It is interesting
to note that setting $m=0$ in the line element one obtains the
vacuum solution ($T_{\mu\nu}=0$). It is also noteworthy that, the
same solution holds for $k=0$ ($m\neq 0$); this latter case was
investigated in \cite{Andr2} where the corresponding solution was
expressed in a different set of variables.

\section{The Hubble parameter} \label{sect4}
The definition of the Hubble parameter in an arbitrary time gauge is
\be \label{hubdef2} \mathrm{H}(t) := \frac{1}{a(t)N(t)} \frac{da
(t)}{dt}. \ee With the use of \eqref{hubdef2} it is an easy task to
evaluate this function for all the solutions obtained in the
previous sections. However if one needs to derive useful conclusions
by comparison to observational data, function $H(t)$ needs to be
expressed in a time coordinate that exhibits a constant gauge
(usually $N=1$). In order to do this, one has to apply the
transformation $t\rightarrow \tau=\int \!\! N dt$, which in most
cases is practically impossible, since a closed form for the
function $t(\tau)$ is not always attainable.

In what follows we evaluate $H(t)$ for all the derived solutions and
wherever possible we give in closed form the transformation that
links $H(t)$ to $H(\tau)$. At first, we consider the cases were
$k=0$:
\begin{itemize}
\item $\gamma=-1$, $\mu\neq \frac{1}{2}$ and $\mu\neq \frac{5}{4}$ \\
    The desired function is
    \be \label{hubble1}
    \mathrm{H}_\mu (t) = \frac{\kappa_3 e^{2 (\mu -1) \sigma } t^{\frac{4 \mu ^2-3 \mu -2}{4 \mu ^2-9 \mu +5}} \left(\kappa_1 \left(2 \mu ^2-3 \mu +1\right) t^{\frac{1}{\mu -1}}+\kappa_3 (4 \mu -5) t^{\frac{2 \mu }{\mu -1}}\right)^{-\frac{2 (\mu -1)}{4 \mu -5}}}{6 \lambda  (\mu -1) \mu }
    \ee
    with $\kappa_3 =6\, \lambda^2 \mu \, (\mu -1)^3  (2 \mu -1) e^{(5-4 \mu ) \sigma}$ being the only fixed constant. The behavior of function \eqref{hubble1} is highly dependent on $\mu$. For example, if we set $\lambda=\sigma=\kappa_1=1$, then for $\mu=\frac{3}{2}$
    \be
    \mathrm{H}_{3/2} = \frac{2 e t^3}{9 t^4+4 e},
    \ee
    while for $\mu=2$
    \be
    \mathrm{H}_{2} = \frac{\sqrt[3]{3} e t^{8/3}}{\left(t \left(36 t^3+e^3\right)\right)^{2/3}}.
    \ee
    It is easy to check that $\mathrm{H}_{3/2}$ is an increasing function of $t$ until the maxima exhibited at $t=\sqrt{2} \sqrt[4]{\frac{e}{3}}$, from then it just decreases until at $t\rightarrow +\infty$, $\mathrm{H}_{3/2}\rightarrow 0$. On the other hand $\mathrm{H}_{2}$ is an increasing function that becomes upper bounded at infinity $\underset{t\rightarrow +\infty}{\lim} \mathrm{H}_{2} = \frac{e}{6 \sqrt[3]{2}}$. By choosing even bigger values of $\mu$, functions that diverge at infinity can be obtained.

    As is evident from \eqref{sollap1} one can not obtain a closed form solution $t(\tau)$ from $\tau(t) =\int\!\! N(t) dt$ for arbitrary values of $\mu$. We only mention that in the case where $\mu=3/2$ the corresponding relation is just $t(\tau)= \frac{e^{\sigma } \tau }{9 \lambda}$. This linear relation between $t$ and $\tau$ means that $\mathrm{H}_{3/2}(t)$ and $\mathrm{H}_{3/2}(\tau)$ exhibit the same functional behavior.
\item $\gamma=-1$ and $\mu= \frac{5}{4}$ \\
    In this case we have obtained the solution space in an arbitrary gauge. But, for the analysis of the Hubble parameter, we have to choose some functional form for $a(t)$. the simplest admissible one (so that the lapse does not become zero) is $a(t)=t$, then the corresponding Hubble parameter becomes
    \be
    \mathrm{H}_{5/4} = \frac{2 \sqrt[3]{2} (\kappa_1 \log (t)-\kappa_2)^{5/6}}{5^{5/6} \kappa_1^{1/6} \lambda^{2/3} t^2}.
    \ee
    It can be seen that $\mathrm{H}_{5/4}$ is an increasing function until the maxima at $t =e^{\frac{5 \kappa_1 +12 \kappa_2}{12 \kappa_1}}$, then it decreases and when $t\rightarrow+\infty$, $\mathrm{H}_{5/4}$ converges to zero. In this case, a closed form expression for $t(\tau)$ cannot be found, since $\tau(t)=\int\!\! N dt$ is expressed in terms of the incomplete Gamma function.
\item $\gamma=-1$ and $\mu= \frac{1}{2}$ \\
    As in the previous case we choose $a(t)=t$. The Hubble parameter is
    \be
    \mathrm{H}_{1/2} = -\frac{2 \kappa_3 e^{-\frac{4 \kappa_3 (\kappa_1+2 \kappa_3)-3 \lambda ^2 t^6}{48 \kappa_3^2}}}{3 \lambda  t^2}.
    \ee
    This is an even function that diverges ($\mathrm{H}_{1/2} \rightarrow - \infty$) for both $t\rightarrow 0$ and $t\rightarrow +\infty$, its maxima are at $t = \pm \frac{2^{2/3} \sqrt[3]{\kappa_3}}{\sqrt[6]{3} \sqrt[3]{\lambda}}$. Again the transformation $t\rightarrow \tau(t)$ cannot be applied.
\item $\gamma\neq -1, \pm \frac{2}{3}$ and $-\frac{4}{3}$ \\
    As we have seen, in this case $\mu=\frac{3(\gamma+1)}{3\gamma+2}$, so we expect $\gamma$ to play a dominant role in the Hubble parameter, with the latter being
    \be
    \begin{split} \label{hubble2}
    \mathrm{H}_{\mu,\gamma} =-(3 \gamma +2) (3 \gamma +4) \kappa_3 \sigma^{\frac{2}{3 \gamma +2}} \left((6 \gamma +4) m+\lambda  t^{3 \gamma +3}\right) \times \\ \frac{\left(5 \kappa_1-(3 \gamma -2) \kappa_3 t \left(2 \left(9 \gamma ^2+18 \gamma +8\right) m+\lambda  t^{3 \gamma +3}\right)\right)^{\frac{2}{3 \gamma -2}}}{90 (\gamma +1) \lambda }
    \end{split}
    \ee
    where $\kappa_3 = \frac{450 (\gamma +1) \lambda  \sigma ^{\frac{3 \gamma -2}{3 \gamma +2}}}{(3 \gamma +2)^2 (3 \gamma +4)}$.

    In the case of stiff matter, $\gamma=1$ and $\mu=\frac{6}{5}$, function \eqref{hubble2} becomes
    \be
    \mathrm{H}_{6/5,1} = \sigma \left(-\left(10 m+\lambda  t^6\right)\right) \left(5 \frac{\kappa_1}{\sigma^{1/5}} -\frac{36}{7} \lambda t \left(70 m+\lambda  t^6\right)\right)^2,
    \ee
    which converges at $t=0$ to $\mathrm{H}_{6/5,1}(0)=-250 \kappa_1^2 m \sigma ^{3/5}$, while for $t\rightarrow \pm \infty$ diverges to $\pm \infty$ depending on the sign of $\lambda$. A positive lambda leads to $\underset{t\rightarrow \pm \infty}{\lim}\mathrm{H}_{6/5,1}=-\infty$.

    For dust, where $\gamma=0$ and $\mu=\frac{3}{2}$, we get
    \be
    \mathrm{H}_{3/2,0} = -\frac{2 \sigma  \left(4 m+\lambda  t^3\right)}{4 \kappa_1 \sigma +720 \lambda  m t+45 \lambda ^2 t^4}.
    \ee
    At zero, $\mathrm{H}_{3/2,0}(0)= -\frac{2 m}{\kappa_1}$, while at infinity $\underset{t\rightarrow \pm \infty}{\lim}\mathrm{H}_{3/2,0} = 0$.

    Finally, if we consider radiation $\gamma=\frac{1}{3}$, $\mu=\frac{4}{3}$ the function
    \be
    \mathrm{H}_{4/3,1/3} = -\frac{3 \sigma  \left(6 m+\lambda  t^4\right)}{5 \left(3 \kappa_1 \sqrt[3]{\sigma }+240 \lambda  m t+8 \lambda ^2 t^5\right)^2}
    \ee
    is obtained. This time $\mathrm{H}_{4/3,1/3}(0) = -\frac{2 m \sqrt[3]{\sigma }}{5 \kappa_1^2}$ and again $\underset{t\rightarrow \pm \infty}{\lim} \mathrm{H}_{4/3,1/3} = 0$.

    Of course, the arbitrariness of $\gamma$ makes the derivation of a general transformation $t(\tau)$ impossible. In the particular subcase $\gamma=0$, the proper time is $\tau = -\frac{45 \lambda  t}{\sigma}$ thus, $\mathrm{H}_{3/2,0}(\tau)$ behaves exactly like $\mathrm{H}_{3/2,0}(t)$.
\item $\gamma = \frac{2}{3}$ \\
    This means that $\mu=\frac{5}{4}$ and it can easily be seen that
    \be
    \mathrm{H}_{5/4,2/3} = -\frac{4 \kappa_3 \sqrt{\sigma } \left(8 m+\lambda  t^5\right) e^{-\frac{5 \lambda  t \left(48 m+\lambda  t^5\right)}{8 \kappa_1^2}}}{25 \lambda },
    \ee
    where $\kappa_3= \frac{25 \lambda }{16 \kappa_1}$. The value of $\mathrm{H}_{5/4,2/3}$ at zero is $-\frac{2 m \sqrt{\sigma }}{\kappa_1}$ and when $t$ goes to plus or minus infinity the function converges to zero.
\item $\gamma= - \frac{4}{3}$ \\
    The corresponding value of $\mu$ is $\frac{1}{2}$ and the Hubble parameter becomes
    \be
    \mathrm{H}_{1/2,-4/3}=\frac{2 \kappa_3 (4 m t-\lambda )}{3 \lambda ^{2/3} t \sqrt[3]{\lambda  \sigma +64 \kappa_3^2 m t-16 \kappa_3^2 \lambda  \log (t)}}.
    \ee
    As it can easily be seen, $\mathrm{H}_{1/2,-4/3}$ is defined for positive $t$ and specifically for those values that satisfy the condition
    \begin{equation*}
    \lambda  \sigma +64 \kappa_3^2 m t-16 \kappa_3^2 \lambda  \log (t)>0.
    \end{equation*}
    However, what can easily be deduced is, that for $t\rightarrow \pm \infty$, $\mathrm{H}_{1/2,-4/3}$ becomes zero.
\end{itemize}
Finally, we are left with the case $k\neq 0$. Under our assumptions,
the sole case that emerged leads to $\mu=\frac{3}{2}$, $\gamma=0$
and \be \label{hubble3} \tensor[_{k}]{\mathrm{H}}{_{3/2,0}} =
-\frac{2 \kappa_3 \left(\lambda  t \left(t^2-18 k\right)+4
m\right)}{9 \lambda \left(2 \kappa_1 \sigma -36 \kappa_3 k \lambda
\sigma  t^2+16 \kappa_3 m \sigma  t+\kappa_3 \lambda  \sigma
t^4-1\right)} \ee with $\kappa_3 =\frac{9 \lambda  \sigma }{4}$.
Since the lapse function is $N=-9 \lambda \sigma$,
$\tensor[_{k}]{\mathrm{H}}{_{3/2,0}}(\tau)$ can be obtained by a
constant scaling of time $t=-\frac{\tau}{9\lambda \sigma}$.
Irrespectively of the choice of $k$, the value of
$\tensor[_{k}]{\mathrm{H}}{_{3/2,0}}(\tau)$ at $\tau=0$ is $-\frac{8
m \sigma }{8 \kappa_1 \sigma -4}$ and for $\tau\rightarrow  +\infty$
the function converges to zero. This of course does not mean that
for different $k$, $\tensor[_{k}]{\mathrm{H}}{_{3/2,0}}$ exhibits
the same behavior. One can easily see, for example that $k=+1$ and
$k=-1$ correspond to quite different functional behaviors, since
that change of sign inside the polynomial of the denominator,
changes the number and the nature of possible extrema throughout the
domain of the function.

\section{Conclusion}
Throughout this paper we exploit the reparametrization invariance of
singular systems, with a Lagrangian of the form \eqref{genlag}.
Autonomous linear integrals of motion in phase space, are generated
by the Killing vector fields of the scaled supermetric
$\bbar{G}_{\alpha\beta}= V G_{\alpha\beta}$, while its proper
conformal Killing vectors give rise to integrals with an explicit
time dependance.

This scheme is implemented in the context of an FLRW space-time in
$f(R)$ gravity minimally coupled with a perfect fluid governed by a
barotropic equation of state. As a first step, we write the
equivalent mechanical system (equivalent in the sense that the field
equations are satisfied whenever the Euler - Lagrange equations
hold). Then we require that the reduced system describes a maximally
symmetric mini-superspace. This imposes conditions upon $f(R)$, as
well as on any other characteristic parameters that enter the
Lagrangian. Since the mini-supermetic is two dimensional there exist
three Killing fields. Thus, the three autonomous integrals of motion
together with one of the infinite rheonomic, are used to completely
integrate the equations of motion. It is interesting to observe that
the latter are not used anywhere in the analysis, since the number
of the integrals define enough independent relations to completely
solve the system (even algebraically in some cases). Some known
solutions are obtained, together with some others that, to our
knowledge, are new to the literature (subsections \ref{new1},
\ref{new2} and the general case in section \ref{new3} with $k,m \neq
0$).

Finally, we conclude with a brief investigation of the resulting
Hubble parameters $H(t)$. It is a common fact in the search of
analytical solutions, that sometimes a time gauge different than
$N=1$ is a more convenient choice to be employed so as to integrate
the field equations. The same holds in our case as well, so most of
the solutions are expressed in time coordinates where the lapse
function is not constant. If one wants to express the Hubble
parameter with respect to the proper time defined by $\tau=\int\!\!
N(t) dt$, the transformation $t=t(\tau)$ must be found. As we
observe from the particular examples, this is a rather difficult
task, even impossible in some cases. However, the gauge freedom of
the lapse function is a too powerful tool, to senselessly restrict
it prior to writing down the equations of motion. Let alone that any
integral of motion that is constant modulo the constraint equation
$H_C \approx 0$, would be impossible to be recovered if $N=1$ was
naively imposed inside the Lagrangian.

\end{document}